\documentstyle[aps,prl,twocolumn,epsf,amssymb]{revtex}
\begin{document}
\epsfverbosetrue
\title{Cluster Monte Carlo: Scaling of Systematic Errors in the 2D Ising Model}
\author{Lev N. Shchur \dag\ and Henk W.J. Bl\"ote \ddag }
\address{
\dag Landau Institute for Theoretical Physics, 117940 GSP-1 Moscow V-334, 
Russia \\
\ddag
Laboratory of Applied Physics, Delft University of Technology,
P.O. Box 5046, 2600 GA Delft, The Netherlands   \\
}
\maketitle
\begin{abstract} 
We present an extensive analysis of systematic deviations in Wolff 
cluster simulations of the critical Ising model, using random numbers
generated by binary shift registers. We investigate how these deviations 
depend on the lattice size, the shift-register length, and the number
of bits correlated by the production rule. They appear to satisfy scaling 
relations.
\end{abstract}

\pacs{02.50.Ng, 02.70.Lq, 05.50.+q, 05.70.Jk, 06.20.Dk}

\draft

The main advantage of cluster Monte Carlo algorithms is that they
suppress critical slowing down~\cite{SW,W}.  For this reason, cluster 
algorithms are being explored extensively~\cite{FL}. This has even led to 
the construction of special-purpose
processors using the  Wolff cluster algorithm~\cite{Tal,TBS}.

The problem of generating random numbers of sufficient quality is known
to be complicated since the first computer experiments~\cite{Knut}.
Many of the widely used algorithms are of the shift-register (SR) 
type~\cite{SR}. These are extremely fast~\cite{KS}, can be implemented 
simply in hardware~\cite{HSSC,Tal2} and produce 'good random numbers' 
with an extremely long period~\cite{SR}.

Ferrenberg et al.~\cite{FLW} found
that the combination of the two most efficient algorithms
(the Wolff cluster algorithm and the shift-register random-number generator)
produced large systematic deviations for the 2D Ising model on a $16\times 16$
lattice (see also~\cite{STS}).
Also random-walk algorithms appeared to be sensitive to effects due to
the random-number generator~\cite{Grass}.

Remarkably, we did not find visible deviations in simulations~\cite{Tal,TS1}
performed on the special-purpose processor with the Wolff algorithm and
a Kirkpatrick-Stoll random-number generator for lattices larger than
$256\times 256$.

Motivated by this paradoxical situation,
we made an extensive analysis of this problem using
SGI workstations at the Delft University and a DEC AXP 4000/620 server at
the Landau Institute.
A total of about two thousands hours of CPU time was spent.

We find several interesting facts.
First, the maximum deviations occur at lattice sizes for which
average Wolff cluster size coincides with the length $p$ of the SR.

Second, the deviations obey scaling laws with respect to $p$: they can 
be collapsed on a single curve. This opens the possibility to
predict the magnitude of the systematic errors in a given
quantity, depending on the lattice size, the shift-register length 
and, to some extent, also on the number of terms in production rule.

Third, the deviations
change sign when we invert the range of the random number: $x \rightarrow
1-x$.  This provides a simple test, in two
runs only, for the presence of systematic errors.

Finally, we introduce a  simple 1D random-walker model explaining how the
correlations in the SR lead to a bias in Monte Carlo results.

As a first step in understanding the results, it is natural to compare
the length scales associated with the Monte Carlo process and the random
generator.  The first characteristic length is the
mean Wolff cluster size $\langle c \rangle$. 
The second characteristic length is the size $p$ of the shift register.
The production rule 
\begin{equation}
x_n=x_{n-p} \oplus x_{n-q}, \label{SR}
\end{equation}
where $\oplus$ is the 'eXclusive OR' operation, leads to three-bit       
correlations over a length $p$. So, it not surprising that the
largest deviations occur at the lattice size $L_{\rm max}$
for which these two lengths coincide. 
Since the mean Wolff cluster size behaves~\cite{W} as the magnetic
susceptibility $\chi$, we expect at criticality that

\begin{equation}
p \propto \chi\propto L^{\gamma/\nu}_{\rm max},
\label{Rel}
\end{equation}
where $\gamma$ and $\nu$ are the 
susceptibility and correlation length exponents respectively.

We performed Wolff simulations of the 2D Ising model at criticality,
using SR with feed-back positions $(p,q)$=(36,11), (89,38),
(127,64) and (250,103) as listed in Ref.~\onlinecite{HBC} and references 
therein.
For each pair $(p,L)$ we took 100 samples of $10^6$ Wolff clusters.
Thus we determined the coefficient in Eq.~(\ref{Rel}):
$p=1.09(1) \; L_{\rm max}^{7/4}$.
Here, and below, the numbers in
parentheses indicate the statistical errors.

The results for the energy deviations 
$\delta E \equiv \langle E/E_{\rm ex}-1 \rangle$ are plotted in 
Fig.~1. The exact results are taken from Ref.~\onlinecite{FF}.
The maximum deviations occur at  $L=$ 7, 12, 15 and 22 respectively,
in agreement with Eq.~(\ref{Rel}).
The inset in Fig.~1 displays the maximum deviations of the energy
$\delta E_{\rm max}$ as a function of the shift-register length.
A fit yields $\delta E_{\rm max} \propto p^{-0.88(2)}$.

The resulting data collapse for the scaled deviations 
$\delta \tilde{E} \equiv  p^{0.88} \delta E$ is 
shown in Fig.~2 versus the scaled system size 
$\tilde{L} \equiv p^{-0.43(5)} L$.
The linear decay on the right obeys 
$\delta \tilde{E} \propto \tilde{L}^{-0.84(4)}$.

If the data for $L>p^{4/7}$ keep following the linear trend in Fig.~2,
the maximum possible deviations can be described by relation

\begin{equation}
\delta E \,\hbox{\lower0.6ex\hbox{$\sim$}\llap{\raise0.6ex\hbox{$<$}}} \,
0.3\; L^{-0.84}\; p^{-0.52}.         \label{dE}
\end{equation}

The results for (127,64) do not fit the curve well. This is no surprise
because shift registers with (p,q) close to powers of 2 are known~\cite{CH}
to produce relatively poor random numbers.

Similarly, we sampled the deviation of the specific heat $C$.
Fig.~3 shows scaled deviations 
$\delta \tilde{C} \equiv p^{0.51(2)} \delta C$ versus
the scaled system size which is the same as for $\delta E$.
For large $\tilde{L}$ this curve behaves as 
$\delta \tilde{C} \propto \tilde{L}^{-0.21(2)}$.
The deviations satisfy
\begin{equation}
-\delta C \,\hbox{\lower0.6ex\hbox{$\sim$}\llap{\raise0.6ex\hbox{$<$}}} \,
0.85 \; L^{-0.21}\; p^{-0.42}   \label{dC}
\end{equation}
but they can also be decribed in terms of a logarithm of $L$ plus a
constant.
 
Fig.~4 shows analogous results for the dimensionless ratio
$Q=\langle m^2 \rangle^2/\langle m^4 \rangle$, which is related to the 
Binder cumulant~\cite{BK}, using 
$\delta \tilde{Q}=\delta Q \; p^{0.60(1)}$
along the vertical scale.
On the right hand side the data behave as 
$\delta \tilde{Q}
\propto \tilde{L}^{-0.45(5)}$. Extrapolation leads to       

\begin{equation}
\delta Q \,\hbox{\lower0.6ex\hbox{$\sim$}\llap{\raise0.6ex\hbox{$<$}}} \,
0.244 \, L^{-0.45}\; p^{-0.41}.     \label{dQ}
\end{equation}
 
In order to explain the origin of the observed deviations,
we present a simple model that captures the essentials of the Wolff
cluster formation process. This model simulates a directed random walk in 
one dimension~\cite{SBH}. 
At discrete times, the walker makes a step to the right with probability 
$\mu$; otherwise the walk ends.
The probability to visit precisely $n$ consecutive nodes is
\begin{equation}
P_{\rm ex}(n)=\mu^{n-1} \; (1-\mu ).                 \label{Pex}
\end{equation}
Now, we simulate this model using a SR random-number generator. 
Each walk starts directly after completion of the preceding one, without 
skipping any random numbers.
First, we use the 'positive' condition $x_n \geq \mu$
for stopping.
Thus, the  random number at start always fulfills the condition
$x_0 \geq \mu$, which ended the preceding walk.

In the simplest case $\mu=1/2$, only the leading bit affects this condition.
As long as the walk proceeds,
the leading bits of the random numbers $x_n$ are zero.
After $p-1$
successful moves, the SR algorithm will produce a number
$x_p$ with the leading bit equal to 1. Thus the walker cannot
visit more than $p$ nodes.

A probabilistically equivalent condition for stopping is
the 'negative' condition  $x_n < 1-\mu$. Then, the leading bit of
$x_0$ must be 0, and for $x_n$ ($n\geq1$) it is 1 until the walk ends. 
The walk cannot stop at the $n=p$, since $x_p \oplus x_{p-q} =0 \oplus 1=1$.

One can calculate the deviation from the exact
value of $P(n)$ at $n=p$, $n=p+q$ and at all linear combinations
of numbers $p$ and $q$. The detailed analysis
will appear elsewhere~\cite{SBH} and here we only mention that
the probability deviation 
$\delta P(n)=(P_{\rm comp}(n)-P_{\rm ex}(n))/P_{\rm ex}(n)$
at $n=p$ for the posititive condition is equal to $(1-\mu)/\mu$.
It is important that a deviation, even at only one point $n=p$,
results in a deviation of the probability function
for the points $n>p$ by $\delta P(n)=(2\mu-1)^2/\mu^4-1 < 0$.
The deviations at the
'resonances' $n=ip+jq$ ($i=1,2,...$ and $-i<j<i$) are positive
and lead to negative deviations of the next points.

Thus, in the case of the positive condition, most of the $\delta P(n)$
are negative. In the case of the negative condition,
$\delta P(n)$ is negative for $n=p\; 2^k\; (k=0,1,2,...)$; this results 
in positive deviations for the following points.

In effect, this replaces the probability $\mu$ by a
new 'effective' probability
$\mu^*$, with $\mu^* >\mu$ for the positive condition and
$\mu^* <\mu$ for the negative condition for most $n>p$.
This provides a qualitative explanation of the deviations in Wolff
simulations. The completion of a Wolff cluster is strongly correlated 
with the value of the random numbers used at that time. Thus, the
three-bit correlations generated by the production rule lead to 
two-bit correlations in the following $p$ random numbers.
In particular when the mean Wolff cluster size is about $p$, 
one may expect serious deviations in the calculated quantities.
 
When one replaces the positive by the negative condition, in effect
the three-bit correlation is inverted. Thus, one expects a change of sign 
of the systematic errors.  We confirmed this for the 2D Ising model.

A simple modification  of the SR (1) is  
to use only one out of every $m$ random numbers generated by the
production rule~\cite{FLW,STS}.
If $m=2^k$, $k=(1,2,...)$ this will lead to the same production rule (1).
For $m=3$ and, as an example, for SR (36,11) the resulting production rule is
(36,24,12,11): a 5-point production rule, i.e.
$x_n=x_{n-36} \oplus x_{n-24} \oplus x_{n-12} \oplus x_{n-11}$.
However, the lowest-order correlations of the resulting random numbers do 
not occur at $n=p=36$, but at $n=48$ because the production rule is 
equivalent with a 4-point one, namely (48,23,11)~\cite{JA}.
The effect due to 4-point correlations appears to dominate over the 5-bit 
effects for $\mu > 1/2$.
The deviations $\delta P(n)$ of Eq.~(6) resemble those for a 3-point 
production rule. But for $\mu$ close to 1 they stand out only at
$n=48\; k$, $k=1,2,..$, and not at linear combinations of other magic
numbers. Their sign is the same
for the positive and negative conditions because the 4-point 
rule correlates an even number of bits).

Next, we investigated these 4-bit effects in the case of Wolff 
simulations, using every third number produced by the rules (36,11) and 
(89,38), and  runs of $10^{9}$ clusters. 
The deviations obey the same scaling laws, but the amplitudes are
about 20 times smaller for each of the quantities 
$E$, $C$ and $Q$, in accordance with the behavior of the 1D model
(see the asterisks in Fig. 2).

For $m=5$ - using only every 5-th number~\cite{FLW} - the effective
production rule correlates 5 bits~\cite{JA}. It leads to 
deviations in 1D model, in particular at $n=p k,\; k=2^i$. They
are less than for the SR of Eq.~(1)~\cite{SBH}.

Very long simulations, using 100 samples $\times\; 10^7$ 
Wolff steps for $m=5$, show that the deviations 
are even smaller than for $m=3$.
Table~\ref{Table1}
displays data for SR (36,11) and (89,38) 
at lattice sizes $L=7$ and $L=12$, respectively,
Similar data are included for $m=3$ and for $m=1$.

So, we propose, in addition, that the systematic deviations
of 2D Ising Wolff simulations are described  by Eqs.~(3-5) for all SR-type
algorithms, but the coefficients should be corrected with a factor of 
roughly $10^{-(m_c-3)}$, where $m_c$ is the number of bits correlated by the
production rule. 

A preliminary analysis~\cite{SB} confirms relation (2) also for the 3D
Ising model. The deviations can also be collapsed on 
universal curves, but the exponents and amplitudes differ from the 2D case.

We conclude that the 1D model provides a useful way for
the analysis of random numbers, in particular for the detection of harmful 
correlations in SR sequences.
The errors in Wolff simulations induced by these correlations satisfy
scaling relations which have a considerable significance for large-scale
Wolff simulations.
For instance, they confirm that in recent simulations~\cite{TS1} of 
the random bond Ising model with lattice sizes $L$ greater than 128, 
the bias due to the (250,103) Kirkpatrick-Stoll rule 
was less than the statistical errors.

As explained above, 3-bit correlations in a
SR production rule lead to 2-bit correlations 
in the first $p$ random numbers used for the construction of a new
Wolff cluster. If the size of the latter grows large in comparison with
$p$, the 2-bit effect will decrease because the amount of correlation
contained in the first $p$ numbers remains finite. Indeed, this is in 
agreement with the power-law decay on the right-hand sides of Figs. 2-4.
Although 3-bit effects seem to be much smaller in the cases ($L,p$) 
investigated by us, there is no reason to believe that they are absent.
Thus, eventually they are expected to end the 
aforementioned power-law decay.

\acknowledgments

We are much indebted to J.R. Heringa for contributing his valuable 
insight in the mathematics of shift-register sequences. 
We acknowledge productive discussions with A. Compagner,
S. Nechaev, V.L. Pokrovsky, W. Selke, Ya.G. Sinai, D. Stauffer
and A.L. Talapov.  L.N.S. thanks the Delft Computational Physics 
Group, where the most of the work has been done, for their kind
hospitality. This work is partially supported by grants
RFBR 93-02-2018, NWO 07-13-210, INTAS-93-211 and ISF MOQ000.

\begin{table}
\caption{Deviations of energy $\delta E$, specific heat $\delta C$ and
ratio $\delta Q$. The statistical error in the last decimal place is
shown between parentheses. We used a shift-register length $p=36$ for
$L=7$ and $p=89$ for $L=12$.
The bias appears to depend strongly on the number $m_c$ of bits
correlated by the production rule.}

\begin{tabular}{|lrrrr|}
\label{Table1}
$L$ & $m_c$ & $\delta E$ & $\delta C$ & $\delta Q$   \\ \hline\hline
7  & 3 &  0.007797 ( 10) & -0.094307 ( 52) &  0.014442  ( 10) \\
7  & 4 & -0.000356 ( 13) &  0.005894 ( 69) & -0.000720  ( 14) \\
7  & 5 & -0.000060 ( 11) &  0.001122 ( 60) & -0.000133  ( 15) \\ \hline
12 & 3 &  0.003345 (  9) & -0.066797 ( 65) &  0.009577  ( 13) \\
12 & 4 & -0.000149 ( 15) &  0.003296 ( 79) & -0.000274  ( 18) \\  
12 & 5 & -0.000003 ( 11) &  0.000136 ( 89) & -0.000009  ( 15) 

\end{tabular} 
\end{table}

\begin{figure*}[h]
\epsfxsize=350pt
\epsffile{fig1.eps}
\caption{Energy deviations $\delta E$ for several SR, namely (36,11):
$\circ$; (89,38): $+$; (127,64): $\Box$; and
(250,103): $\blacktriangle$. 
The inset shows the maximum value of $\delta E$ as a function of $p$.} 
\label{Fig1}
\end{figure*}

\begin{figure*}[h]
\epsfxsize=350pt
\epsffile{fig2.eps}
\caption{Scaled deviation of the energy $\delta E$ versus the scaled 
system size, for several SR. 
The symbols are defined in the caption to Fig.~1.} 
\label{Fig2}
\end{figure*}

\begin{figure*}[h]
\epsfxsize=350pt
\epsffile{fig3.eps}
\caption{Scaled deviation of specific heat $\delta C$ versus the scaled
system size, for several SR. 
The symbols are defined in the caption to Fig.~1.}
\label{Fig3}
\end{figure*}

\begin{figure*}[h]
\epsfxsize=350pt
\epsffile{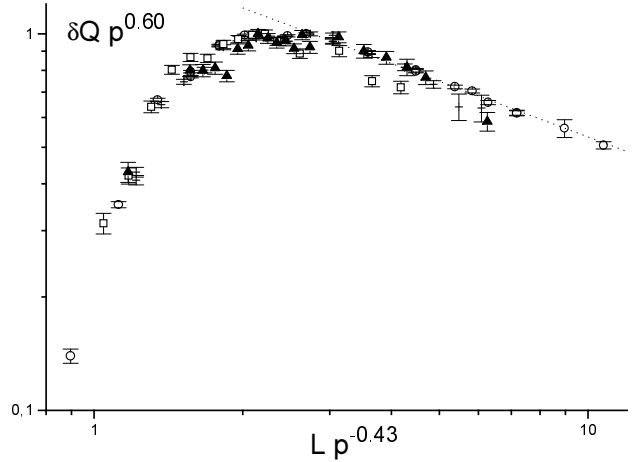}
\caption{Scaled deviation of dimensionless ratio $\delta Q$ versus the scaled
system size, for several SR. 
The symbols are defined in the caption to Fig.~1.}
\label{Fig4}
\end{figure*}


\begin{references}
\bibitem{SW} R.H. Swendsen and J.-S. Wang, 
        Phys. Rev. Lett. {\bf 58}, 86 (1987).
\bibitem{W} U. Wolff, Phys. Rev. Lett. {\bf 62}, 361 (1989).
\bibitem{FL} D. Stauffer, ed., {\it Annual Reviews of Comp. Physics I}
             (World Scientific, 1994).
\bibitem{Tal} A.L. Talapov, L.N. Shchur, V.B. Andreichenko, and
        Vl.S. Dotsenko,  Mod. Phys. Lett. B, {\bf 6}, 1111 (1992).
\bibitem{TBS} A.L. Talapov, H.W.J. Bl\"ote and L.N. Shchur,
        Pis'ma v ZhETF {\bf 62}, 157 (1995);
        JETP Lett. {\bf 62}, 174 (1995).
\bibitem{Knut} D.E. Knuth, {\it The Art of Computer Programming}
              (Addison-Wesley, 1981), Vol. 2.
\bibitem{SR} S.W. Golomb, {\it Shift Register Sequences}
              (Holden-Day, San-Francisco, 1967).
\bibitem{KS} S. Kirkpatrick and E. Stoll, J. Comp. Phys. {\bf 40}, 517 (1981).
\bibitem{HSSC} A. Hoogland, J. Spaa, B. Selman, and A. Compagner,
        J. Comp. Phys. {\bf 51}, 250 (1983).
\bibitem{Tal2} A.L. Talapov, V.B. Andreichenko, Vl. S. Dotsenko, 
        and L.N. Shchur, JETP Lett. {\bf 51}, 182 (1990).
\bibitem{FLW} A.M. Ferrenberg, D.P. Landau, and Y.J. Wong, Phys. Rev. Lett.
              {\bf 69}, 3382 (1992).
\bibitem{STS} W. Selke, A.L. Talapov and L.N. Shchur, JETP Lett. {\bf 58},
              665 (1993);
             K. Kankaala, T. Ala-Nissil\"a and I. Vattulainen,
             Phys. Rev. E {\bf 48}, R4211 (1993);
             P.D. Coddington, Int. J. Mod. Phys. {\bf C 5}, 547 (1994).
\bibitem{Grass} P. Grassberger, Phys. Lett. A {\bf 181}, 43 (1993).
\bibitem{TS1} A.L. Talapov and L.N. Shchur, Europhys. Lett. {\bf 27},
              193 (1994);
              A.L. Talapov and L.N. Shchur, J. Phys. C: Condens. Matter
              {\bf 6}, 8295 (1994).
\bibitem{HBC} J.R. Heringa, H.W.J. Bl\"ote and A. Compagner,
              Int. J. Mod. Phys. C {\bf 3}, 561 (1992).
\bibitem{FF} A.E. Ferdinand and M.E. Fisher, Phys. Rev. {\bf 185}, 832 (1969).
\bibitem{CH} A. Compagner and A. Hoogland, J. Comp. Phys. {\bf 71}, 391 (1987).
\bibitem{BK} K. Binder, Z. Phys. B {\bf 43}, 119 (1981).
\bibitem{SBH} L.N. Shchur, H.W.J. Bl\"ote and J.R. Heringa (unpublished).
\bibitem{JA} J. Heringa and A. Compagner, private communication (1996).
\bibitem{SB} L.N. Shchur and H.W.J. Bl\"ote, unpublished (1996).
\end{references}
\end{document}